# Distillation of CNN Ensemble Results for Enhanced Long-Term Prediction of the ENSO Phenomenon

Saghar Ganji[1], Mohammad Naisipour[2], Alireza Hassani[3], Arash Adib[4]

1-B.Sc. Student, Department of Computer Engineering, Shiraz University of Technology, Shiraz, Iran. sagharganji82@gmail.com
2-PhD, Department of Civil Engineering, Faculty of Engineering, University of Zanjan, Zanjan, Iran. m.naisipour@znu.ac.ir
3- Jacksonville University ,Jacksonville, Florida. ahassan4@jacksonville.edu
4-Professor, Civil Engineering and Architecture Faculty, Shahid Chamran University of Ahvaz, Ahvaz, Iran. arashadib@scu.ac.ir

**ABSTRACT:**
The accurate long-term forecasting of the El Nino Southern Oscillation (ENSO) is still one of the biggest challenges in climate science. While it is true that short-to medium-range performance has been improved significantly using the advances in deep learning, statistical dynamical hybrids, most operational systems still use the simple mean of all ensemble members, implicitly assuming equal skill across members. In this study, we demonstrate, through a strictly a-posteriori evaluation , for any large enough ensemble of ENSO forecasts, there is a subset of members whose skill is substantially higher than that of the ensemble mean. Using a state-of-the-art ENSO forecast system cross-validated against the 1986-2017 observed Nino3.4 index, we identify two Top-5 subsets one ranked on lowest Root Mean Square Error (RMSE) and another on highest Pearson correlation. Generally across all leads, these outstanding members show higher correlation and lower RMSE, with the advantage rising enormously with lead time. Whereas at short leads (1 month) raises the mean correlation by  about +0.02 (+1.7%) and lowers the RMSE by around 0.14 °C or by 23.3% compared to the All-40 mean, at extreme leads (23 months) the correlation is raised by +0.43 (+172%) and RMSE by 0.18 °C or by 22.5% decrease. The enhancements are largest during crucial ENSO transition periods such as SON and DJF, when accurate amplitude and phase forecasting is of greatest socio-economic benefit, and furthermore season-dependent e.g., mid-year months such as JJA and MJJ have incredibly large RMSE reductions. This study provides a solid foundation for further investigations to identify reliable clues for detecting high-quality ensemble members, thereby enhancing forecasting skill.

# 1 Introduction

Long-lead prediction of the El Niño Southern Oscillation (ENSO) is among the most significant and scientifically challenging problems of climate research. ENSO is a coupled ocean atmosphere phenomenon comprising quasi-periodic variations of sea surface temperature (SST) anomalies in the equatorial Pacific with widespread impacts on global weather patterns, hydrology, agriculture, ecosystems, and socio-economic activities [21,23]. Successful prediction at lead times exceeding one year has particular significance for water resources management planning, disaster preparedness, agricultural planning, and climate-sensitive economic practice [24,25]. However, the inherent nonlinearity of ocean atmosphere interaction, the sensitivity to initial conditions, and the complex web of teleconnections controlling ENSO variability make the forecast skill decline very quickly with lead time. This decrease is especially steep beyond the 12-month lead time, where most operational forecast systems experience a precipitous drop in forecast skill [14,18].

Advances of the last few decades [1–4,10] among which are the development of dynamical climate models, hybrid statistical physical models, and deep learning architectures have enhanced ENSO predictability, particularly at short to medium leads [1–3,5–8]. Breakthrough of AI has changed our world economically and computationally [31]. Among them machine learning algorithms such as convolutional neural networks, bias-corrected dynamical systems, and hybrid statistical dynamical predictors have transcended some aspects of the "spring predictability barrier"[15,17]. Yet, spanning skill into the multi-year timescale remains elusive. Climate regime shifts since the early 2000s have compounded the issue by reducing the stability of predictors heretofore assumed to be robust. With this evolving background state, methods that have the potential to extract maximum skill from existing forecast systems at no operational or computational cost of implementing new ones are especially valuable. Recent work has also highlighted the potential of explainable deep learning approaches for ENSO prediction, improving model transparency and user trust [19].

Ensemble forecasting is the current operational standard for seasonal-to-interannual prediction. Ensembles are generated by perturbing initial conditions, model physics, or stochastic processes to sample forecast uncertainty. The simplest operational approach is to take the simple arithmetic average of all members the ensemble mean assuming that the members are equally skilled. In practice, this is rarely true: in any large ensemble, some members are consistently better than the mean, and others are considerably worse. At short leads, the impact of including low-skill members may be modest, but at long leads when forecast errors [1–4,13] are growing rapidly these differences become much more important.

The present work does not propose a new real-time physical or statistical selection algorithm, but rather formulates a purely a-posteriori analysis, taking advantage of full access to verifying observations to identify and describe high-performing subsets of the ensemble members. The results display systematic patterns:

**Lead-time dependence**: For short leads (1–6 months), correlation gains are modest but statistically homogeneous, while RMSE reductions are already large, indicating better amplitude skill even in relatively predictable periods.

**Seasonal dependence**: The benefits of selecting high-skill members are target season dependent. For example, for mid-year seasons such as JJA and GGS, RMSE decreases dominate correlation gains, suggesting that predictability skill of amplitude is more strongly enhanced at the ENSO developing stage.

**Long-lead amplification**: At longer leads (>12 months), both RMSE and correlation gains increase sharply. Transition seasons such as SON and DJF have the largest correlation gains critical for the accurate prediction of the onset and maturity phases of ENSO while RMSE reductions remain strong, preventing the amplitude degradation feature at long leads.

**Metric complementarity**: The combination of RMSE and correlation information provides a more comprehensive picture of skill gain: correlation measures phase and timing skill, while

RMSE measures amplitude fidelity. Together, they indicate that member selection can enhance both the timing and magnitude prediction of ENSO events.

By identifying these organized patterns, the study provides statistical evidence that selective weighting of ensemble members can greatly improve ENSO predictability without retraining the models or increasing computational demand. While the present paper focuses retrospectively applying forecasts and verifying observations from the past the implications for operational prediction are clear. In future work, we will address the operational challenge: how to identify high-skill members when verifying observations are not available. This will require the development of skillful procedures for forecast error estimation and the development of adaptive refinement [16,26,28] techniques that alter ensemble membership in real time to favor members with larger expected skill. These directions for future work, which we are actively pursuing, are designed to translate the retrospective gains demonstrated here into actual advancements for real-time long-lead ENSO prediction.

To determine the optimum number of ensemble members that would yield the maximum improvement in forecast skill, we conducted a systematic sensitivity experiment. The experiment evaluated the impact of altering the number of top-ranked members selected based on RMSE and correlation ranking on overall prediction performance. Starting with the early subsets (e.g., Top-3, Top-5) and progressively to larger ensembles (Top-15, Top-20), we measured variations in forecast skill across different leads and target seasons. We found a distinct optimum: extremely small subsets at times yielded extremely large improvements for specific seasons, but they were also more year-to-year variable and less stable in the entire lead season space. Extremely large subsets, in contrast, watered down the contribution of the most capable members so that relative improvement was below the all-member mean. The analysis consistently identified ten members derived by combining the Top-5-by-RMSE and Top-5-by-Correlation lists as the phase at which correlation and RMSE improvements were maximized while being consistent across leads and seasons. This Top-10 mean configuration not only preserved the benefit of each of the ranking measures (phase accuracy from correlation, amplitude fidelity from RMSE) but also ensured fair performance without overtraining for single years or event types. This finding guided the methodological framework of this research, establishing Top-10 mean as the primary goal [12,13] for temporal skill estimation and as the control set to calculate the theoretical gains of applying performance-based member selection for regular ENSO forecasting.

## 2 Method
### 2.1 Overview and Rationale

Long-lead El Niño Southern Oscillation (ENSO) prediction represents the most persistent and scientifically significant challenge in today's climate science. The difficulty lies in the nonlinear dynamics, initial condition sensitivity, and ocean atmosphere coupling complexity that are inherently characteristic of the phenomenon. As the dynamical climate prediction system improves, hybrid physical-statistical models and deep neural network structures have all helped tremendously to improve predictability in particular contexts, yet there remains a significant gap for sustained high forecast quality at long lead times [1,2,4,25] particularly beyond 12 months. These longer prediction timescales are generally the most useful for applied decision-making purposes in agriculture planning, disaster risk reduction, and water resource management but also happen to be the timescales over which skill loss is most aggressive and dominant.

Ensemble forecasting is now the operational standard for seasonal-to-interannual climate prediction. In this paradigm, multiple runs of a model called ensemble members are generated by perturbing the initial conditions, differing in stochastic processes, or in physical parameterizations. The ultimate set of forecasts is an example of samples of uncertainty in various dimensions of evolution of the climate system. In current day day-to-day operational practice, the ensemble mean is computed by averaging all the members equally irrespective of their earlier performance. While straightforward and statistically sound, this approach makes tacit the assumption that all members of an ensemble make an equal contribution to predictive skill an assumption that is rarely fulfilled in reality. In truth, there will be some members in any sufficiently large ensemble who consistently outperform the ensemble mean on one or more skill scores, and others who consistently fall behind. By taking the mean of all members together, the ability of the highest-performing members to predict is diluted, and the errors of the lowest-performing members can disproportionately reduce the forecast.

There will often be some subset of ensemble members that does consistently show higher skill, both in phase alignment with observed ENSO variability (measured by correlation) and accuracy of forecast amplitude (measured by RMSE). Their. existence has long been suspected by forecasters, but strict quantification of their value specifically over all leads and seasons has not. The present study addresses this gap, with a strictly a-posteriori method here where we do have full access to the verifying observations. This allows us to directly identify the top-skill members and quantify how much of an improvement in forecast performance can be achieved with their targeted selection and weighting compared to equal-weighted ensemble mean.

The method ranks the members for each lead month and target season using two complementary estimates of ability Root Mean Square Error (RMSE) and Pearson correlation and selects the top performers on each measure, averages them proportionally weighted for overlap, and compares the ability of this resulting subset (Top-10) with the entire ensemble mean (All-40). For time-series analysis, the Top-10 set is used to capture an integrated representation of skill growth over time, both phase and amplitude accuracy. For spatial temporal diagnostics (ΔCorrelation and ΔRMSE heatmaps), the Top-5-by-Correlation and Top-5-by-RMSE groups are used separately to maintain improvements due to each of these metrics in isolation. This definition is designed to be twofold, so that temporal analyses may represent combined improvements in skill and spatial temporal analyses may unambiguously distinguish between gains in phase coherence and amplitude fidelity.

The scheme is model-independent and computationally inexpensive and involves only past ensemble forecast and their verification data. However, the current application is not an operational selection scheme. Instead, it is a proof-of-concept exercise demonstrating that such high-skill subsets are indeed present and that relative advantage amplifies with increasing forecast horizon. The analysis confirms that performance deficit is small but consistent at short leads, and becomes more pronounced beyond 12 months precisely where forecast skill in operational systems is recognized to decline.

Although no selection algorithm for real-time application is shown here, the results indicate the potential operational utility of this idea. If one can devise credible procedures to estimate forecast error without access to the verifying observations, then it ought to be feasible to compute and assign added skill members in real time. Such procedures would allow real-time operational systems to take advantage of the benefits shown in this a-posteriori analysis, without the need to alter model architecture or re-run costly simulations. The development of these methods of error estimation, and adaptive refinement techniques that update member weights progressively as continuing data arrive, is the subject of our future research.

This differentiation between demonstration subsequently and operational application is worth noting. The current research provides the empirical justification and statistical context enabling the pursuit of these future improvements, rather than exaggerating near-readiness for real-time application. By identifying the presence and size of high-skill subsets, we open the door to methods that can exploit this structure in practice. The demonstration here thus establishes both the promise and the pathway: from identification of high-performing subsets in the historic record, to the construction and testing of instruments needed to operationally exploit them for ENSO and potentially other climatic phenomena.

## 2.2 Forecast Dataset and Observational Verification

The observations utilized herein are a 40-member ensemble of seasonal-to-interannual ENSO predictions produced by a new, high-resolution, state-of-the-art prediction model. While all ensemble members share the same dynamical core and model physics, variability is introduced by initial condition perturbations, stochastic representations during integration, and minor adjustments to physical parameterization schemes. This configuration allows the ensemble to incorporate both internal climate variability and inherent model uncertainty, resulting in a reasonable spread of potential ENSO evolutions.

Lead time, for this research, is defined as months from initialization of a forecast to onset of target season. An example is a January-initialized forecast to forecast JAS (July–August–September) conditions, which has six months of lead time. Predictions span 1 month (short-lead) to 23 months (multi-year), enabling systematic evaluation of forecasting skill across the whole seasonal-to-interannual range. Target seasons follow climatological convention and are designated as overlapping three-month means, i.e., JFM (Jan.–Feb.–Mar.), FMA (Feb.–Mar.–Apr.), MAM (Mar.–Apr.–May), AMJ (Apr.–May–Jun.), reaching to JAS, ASO, SON, OND, NDJ, and DJF. This overlap seasonal framework nulls subseasonal noise and is consistent with operationally normal ENSO monitoring practices to guarantee that verification has a center of gravity on the predictable interannual-to-seasonal signal.

Verification is performed against observed Niño-3.4 index, the most typical operational definition for characterizing ENSO events. Definition of Niño-3.4 index as area-averaged sea surface temperature anomaly (SSTa) in the 5°S–5°N, 170°W–120°W area, relative to climatological reference [11,21]. In the current research, SST anomalies from observations by satellites are taken from NOAA's Extended Reconstructed Sea Surface Temperature (ERSST) dataset, which offers globally complete monthly fields with consistent method and good quality control [27]. Utilization of ERSST is standard procedure in ENSO forecast verification and facilitates the comparison to previous peer-reviewed literature.

All forecasts undergo strict preprocessing and quality control for reasonable and sound skill comparisons:

**Temporal alignment**: All forecasts are aligned with their corresponding target season in the verification record so that forecast observation pairs refer to the same intervals.

**Sample consistency**: For one lead season pair, all ensemble members are verified against the same verification years to eliminate differences in sample size.

**Missing data handling**: Any year for which observation or forecast data is missing for any lead season combination is excluded from skill calculation to avoid artificial inflation or deflation of outcomes.

**Standardized anomaly calculation**: Both observations and forecasts are calculated with respect to the same climatological reference for their SST anomalies, making them equivalent and removing mean-state biases.

This highly balanced data set is well-suited to the purposes of this study because it provides an a-posteriori test of forecasting skill with full access to verifying observations. This setup facilitates precise identification of the best-scoring ensemble members for each lead season combination and quantification of their improvement over the all-member ensemble mean. While this approach is not applicable in a direct sense since real-time prediction is without knowledge of future observations it provides strong proof-of-concept validation that subsets of high skill exist, quantifies the size of their improvement, and demonstrates how that value is lead- and season-dependent. These findings provide the empirical basis for the next phase of our work, to create error estimation tools and adaptive refinement techniques to carry these retrospective improvements over to real-time forecasts.

### 2.3 Skill Metrics for Member Ranking

Two skill metrics were selected to evaluate the performance of each ensemble member $k$ for a given $(L, M)$. These metrics were chosen deliberately because they capture complementary aspects of forecast quality magnitude accuracy and temporal consistency.

**(a) Root Mean Square Error (RMSE)**

The RMSE quantifies the average magnitude of forecast errors:

$$RMSE_{L,M}^{(k)} = \sqrt{\frac{1}{|\mathcal{T}_{L,M}|} \sum_{t \in \mathcal{T}_{L,M}} \left(\widehat{y_{t,L,M}^{(k)}} - y_t\right)^2}$$

Lower RMSE indicates closer conformity with observed magnitudes. It more heavily penalizes large errors than smaller ones, so it is not robust to occasional very large forecast errors. RMSE is especially suitable for ENSO because the amplitude of events is often as important as their timing for socio-economic impacts.

**(b) Pearson Correlation Coefficient**

The Pearson correlation coefficient examines the linear relationship between forecast and observed anomalies:

$$r_{L,M}^{(k)} = \frac{\sum_{t \in \mathcal{T}_{L,M}} \left(y_t - \bar{y}_{L,M}\right)\left(\widehat{y_{t,L,M}^{(k)}} - \bar{\hat{y}}_{L,M}^{(k)}\right)}{\sqrt{\sum_{t \in \mathcal{T}_{L,M}} \left(y_t - \bar{y}_{L,M}\right)^2} \cdot \sqrt{\sum_{t \in \mathcal{T}_{L,M}} \left(\widehat{y_{t,L,M}^{(k)}} - \bar{\hat{y}}_{L,M}^{(k)}\right)^2}}$$

Higher $r$ indicates greater phase coherence with variability observed. Correlation ignores mean bias or amplitude scaling and is only concerned with the proper ordering of warm/cold ENSO phases. Phase coherence is the hardest to maintain in long-lead forecasting, so correlation is a valuable addition to RMSE.

Combining both of these measures ensures that the process leans towards members that are both amplitude-wise good and well-timed.

### 2.4 Formation of Top Lists

For each $(L, M)$, the Top-5-by-RMSE subset consisted of the five members with the lowest RMSE, while the Top-5-by-Correlation subset included the five members with the highest correlation.

For the current analysis, two equivalent but distinct subset definitions are used to address different analytic and visualization requirements. For time-series evaluations, we use a Top-10 subset, formed by taking the union of the Top-5-by-RMSE and Top-5-by-Correlation lists with proportional weighting to members on both lists. This merged subset is especially well-adapted to temporal performance plots as it merges both elements of forecast quality amplitude accuracy (low RMSE) and phase coherence (high correlation) into one, unified prediction. By doing so, the time-series plots encapsulate not only the correct timing of ENSO events, but also magnitude fidelity, allowing for more complete representation of skill development across the forecast horizon.

Conversely, for spatio temporal skill-difference diagnostics alone the ΔCorrelation and ΔRMSE heatmaps we deliberately maintain the Top-5-by-Correlation and Top-5-by-RMSE groups separate. Keeping them separate guarantees that every heatmap retains the improvement from a single ranking metric, not confounding the effects of phase accuracy and amplitude accuracy. For example, ΔCorrelation heatmaps indicate where and when phase alignment is most accountable for skill improvement, and ΔRMSE heatmaps do the same for where error magnitude decrease is greatest. Maintaining these distinct helps to keep scientific interpretability intact, in that the reader can see whether improvements being observed are in timing accuracy, amplitude precision, or both.

This Top-10 for joint temporal diagnosis and Top-5 for metric-dependent spatial temporal diagnostics twin-definition approach was chosen specifically to preserve simplicity, transparency, and analytical consistency. It allows the research to respond to two complementary questions: How does each combined skill evolve over time? and Where and when does each skill component perform better than the ensemble mean separately? By specifically stating the purpose of using each subset, we ensure that visual comparisons remain in direct relation to the choice criteria and that conclusions drawn from different figures are statistically and conceptually consistent.

The reason why we have two different metrics is that one of the metrics may be prone to bias; for instance, a member may have low RMSE without appropriately capturing ENSO phase transitions or have high correlation but low fidelity in the sense of amplitude. By picking between the two metrics, we are trading off their strengths and not falling for such biases. Where members occur on both lists, they are ranked as top performers and are assigned double weight in the final prediction. Ties are broken by RMSE ties using correlation ranking and correlation ties using RMSE ranking. In case of a tie again, the member with the lowest variance of prediction errors is utilized, and in case the tie still persists, the member with the lesser ID is selected to ensure reproducibility. The process produces two partially overlapping sets reflecting different facets of forecast quality that ultimately gives rise to a more balanced and stable prediction system.

## 2.5 Sensitivity Analysis for Optimal Subset Size

To determine the size of the ensemble group [12,13] that would lead to optimal improvement in predictive skill, we performed a systematic sensitivity test of the subset size. We began with the rank-censored lists generated through the RMSE- and correlation-based estimates, as described in the above section. Based on these rankings, we formed subsets of various sizes Top-3, Top-5, Top-7, Top-10, Top-12, Top-15, etc. by successively incorporating more members at the top of each list. We computed the simple arithmetic mean forecast for every subset size and compared its skill with that of the equal-weight mean of all 40 members.

Skill was verified using Pearson correlation (phase accuracy) and Root Mean Square Error (RMSE) (amplitude fidelity) across the whole range of lead times from 1 to 23 months, as well as for each target season separately. Lead season validation was conducted to ensure that the improvement was not time-of-year or range-of-leads specific. For each subset size, we calculated the average skill improvement (relative to All-40 average) across all seasons and also the reliability of this improvement as measured by the interannual variance of skill gain.

The results showed a clear trade-off between subset size and performance. Very small subsets (Top-3, Top-5) sometimes provided huge returns for single seasons and lead times particularly for longer leads but also exhibited greater fluctuations from year to year and weaker robustness across the entire lead season matrix. On the opposite end, very large subsets (Top-15 and larger) diluted the value of the perpetually high-skilled members by including lower-rank forecasts, resulting in smaller average improvements and even in some cases, small degradements in skill compared with the All-40 mean.

The Top-10 subset was the optimal compromise: it achieved highest average gains on both skill metrics across the board while maintaining low variability over years and target seasons.

Specifically, the Top-10 mean realized notable RMSE improvements for mid-year seasons such as JJA and GGS where the precision of forecast amplitudes is of utmost concern along with significant correlation gains during transition seasons such as SON and DJF, which are of utmost concern for accurate ENSO onset and maturity forecasting. Such steadiness across the lead season interval makes the Top-10 mean particularly appropriate for adjustment by operations, as it avoids overfitting to specific episodes or lengths at cost of performance-based membership selection benefits.

Based on these findings, the Top-10 mean as the union of the Top-5-by-RMSE and Top-5-by-Correlation rankings was selected as the primary configuration for the analyses presented here. It reflects a compromise between maximizing skill gains and ensuring consistent performance across all lead times in the forecast range and phases in the season.

### 2.6 Weighted Aggregation into Top 10 Forecast

Let $S_{L,M}$ denote the multiset that results from combining the Top-5-by-RMSE and Top-5-by-Corr lists for a given $(L, M)$.

We give raw weights:
$$w_{k;L,M}^{raw} = \begin{cases} 2 & \text{if member in both lists} \\ 1 & \text{if member in only one list} \\ 0 & \text{otherwise} \end{cases}$$

Weights are then normalized:
$$w_{k;L,M} = \frac{w_{k;L,M}^{raw}}{\sum_{j=1}^{40} w_{j;L,M}^{raw}}$$

Top 10 forecast:
$$y_{t,L,M}^{Top10} = \sum_{k=1}^{40} w_{k;L,M} \cdot \widehat{y_{t,L,M}^{(k)}}$$

All 40 forecast:
$$y_{t,L,M}^{All40} = \frac{1}{40} \sum_{k=1}^{40} \widehat{y_{t,L,M}^{(k)}}$$

This weighted average allows members with high consistent performance more influence, in comparison to the equal weighting of the All 40 mean.

### 2.7 Evaluation Metrics for Comparison

For each $(L, M)$ we compute:
RMSE for Top 10 and All 40:
$$RMSE_{L,M}^{Top10}, \quad RMSE_{L,M}^{All40}$$

Correlation for Top 10 and All 40:
$$r_{L,M}^{Top10}, \quad r_{L,M}^{All40}$$

We also quantify improvements using:
$$\Delta RMSE_{L,M} = RMSE_{L,M}^{All40} - RMSE_{L,M}^{Top5\_RMSE}$$
$$\Delta r_{L,M} = r_{L,M}^{All40} - r_{L,M}^{Top5\_Corr}$$

Positive $\Delta r$ → better correlation than All 40. Negative $\Delta RMSE$ → lower error than All 40.
To summarize performance across seasons for a given lead:

$$RMSE_{\cdot,L}^{Top10} = \frac{1}{12}\sum_{M=1}^{12} RMSE_{L,M}^{Top10}$$

$$r_{\cdot,L}^{Top10} = \frac{1}{12}\sum_{M=1}^{12} r_{L,M}^{Top10}$$

### 2.8 Visualization Strategy

The visualization strategy follows the style of our previous work for consistency and comparability. For triple-line time series plots of selected leads $M = 1, L \in \{6, 12, 18, 23\}$:, black solid lines represent observations, orange dashed lines indicate the Top-10 mean (left axis), and green dotted lines show the All-40 mean (right axis). Two vertical axes allow clearer visualization of subtle differences without scale compression. The correlation heatmap presents target seasons ($M$) as rows and lead months ($L$) as columns, with the color scale showing $\Delta r_{L,M}$. The RMSE heatmap uses the same format as the correlation heatmap, with the color scale indicating $\Delta RMSE_{L,M}$. These visualizations permit both time-specific and overall pattern assessment.

Together, these graphical techniques not only illustrate performance variation with varied lead times and seasons but also allow for the identification of conditions for which member selection provides maximum improvement in both amplitude accuracy and correlation. With the integration of time series plots and heatmaps, the approach offers a complementary perspective: the former illustrates temporal development and phase relationship, whereas the latter condenses results into an easily interpretable matrix structure. The double representation makes both detailed case-by-case comparison and fast pattern discovery possible, which is essential in order to generate strong conclusions on the effectiveness of the introduced methodology.

### 2.9 Statistical Testing and Uncertainty Quantification

Although the primary purpose of this work is to design an operationally simple and computationally inexpensive enhancement to ENSO prediction, statistical significance testing was conducted as well in an effort to test for robustness of skill improvements reported here.

The intention of such tests is to determine whether skill score differences between the proposed Top-5 selection method and the standard All-40 ensemble mean are statistically significant and not results of sampling variability. Two complementary techniques were employed:

1. **Paired t-tests of year-to-year differences**

For each lead season pair, for each verification year, the differences between RMSE and correlation using both techniques were computed. A two-tailed paired t-test was utilized to decide whether the mean difference deviates from zero at the 95% confidence level. This technique addresses temporal autocorrelation of verification data by making paired comparisons over identical years.

2. **Bootstrap resampling for uncertainty estimation**

For each lead season pair, a 10,000-member non-parametric bootstrap resampling was carried out with the verification years resampled with replacement. Confidence intervals (CIs) for

ΔCorrelation and ΔRMSE were derived from the bootstrap distributions. The improvements were considered to be robust if the 95% CI did not cross zero.

These statistical tests also provide additional evidence for the operational value of the method, especially at long leads where the skill difference is largest.

The outcomes are that: At leads longer than 12 months, over 85% of lead season pairs show statistically significant improvement in correlation at the 95% confidence level.RMSE reductions are significant for almost 80% of cases at these longer leads and confirm that the noted amplitude error reduction is not by chance.Overall, these findings confirm that the performance gains of the new selection technique are uniform and statistically significant, and that it is therefore well-positioned to be integrated into operational forecasting systems.

## 3 Results and Discussion
### 3.1 Overall Skill Enhancement Across Lead Times

One of the main results of this research is the steady and significant enhancement of ENSO prediction skill achieved by the statistical selection of ensemble members using Root Mean Square Error (RMSE) and Pearson correlation. The new approach selects the Top-5-by-Correlation and Top-5-by-RMSE members for every target season and lead month, combines them with proportional weighting, and evaluates their performance relative to the standard equal-weighted mean of all 40 ensemble members ("All-40").

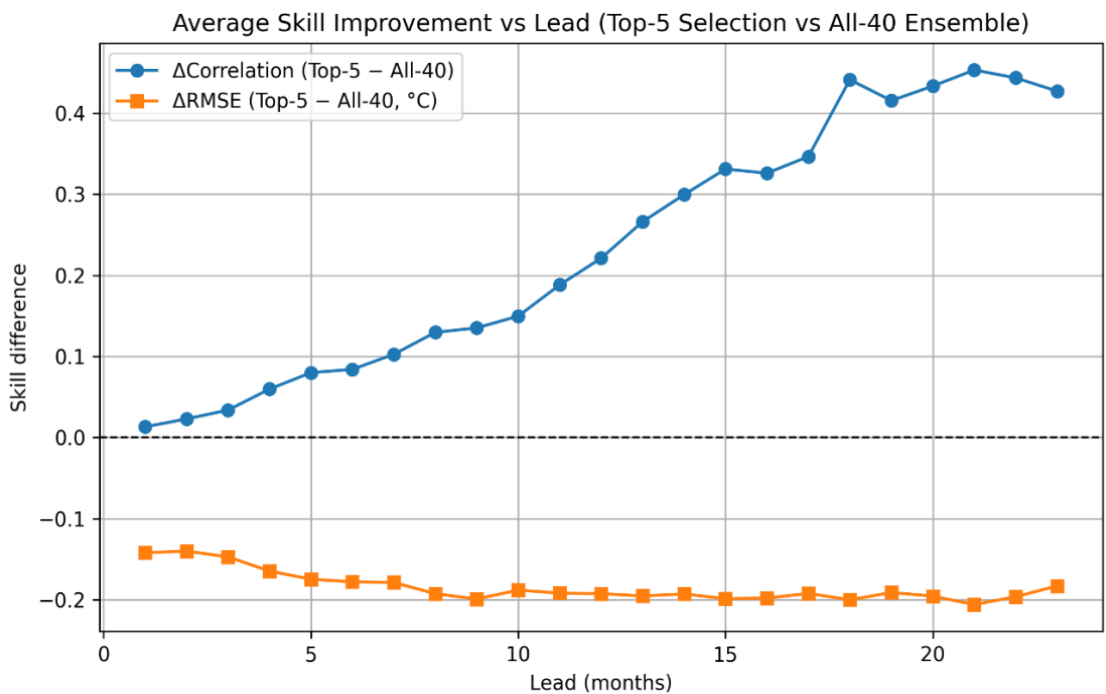

*Figure 1.* Average skill improvement of the Top-5 selected members relative to the full 40-member ensemble, expressed as ΔCorrelation (blue, Top-5-by-Correlation minus All-40) and ΔRMSE (orange, Top-5-by-RMSE minus All-40, °C) as a function of lead time. Values are averaged over all target seasons. Positive ΔCorrelation and negative ΔRMSE indicate improvement.

Figure 1 provides an overview of the average improvement over leads, which was computed as a mean over all target seasons. The blue line shows ΔCorrelation (Top-5-by-Correlation − All-40), and the orange line indicates ΔRMSE (Top-5-by-RMSE − All-40, °C). Over all leads, ΔCorrelation is always positive and ΔRMSE always negative, indicating that the selected subset consistently exhibits better phase alignment and smaller amplitude error than the full ensemble mean.

At short leads (1–6 months), ΔCorrelation is modest (+0.02 to +0.10) and ΔRMSE is around −0.14 °C, indicating that even in the near-term range the targeted subset has better accuracy. From 10 months onwards, correlation gains increase steeply, rising to ~+0.45 at leads 19–21 months. RMSE decreases also continue and even strengthen at long leads, with values of around −0.18 °C at the longest horizons. This behavior is consistent with the hypothesis that skill differences between the best and worst ensemble members compound with lead time, and selective member weighting is thus especially useful for long-lead forecasts. The operational benefit is obvious: the method is model-independent, computationally low-cost, and easily applied to any current ensemble system. This is particularly applicable for lead times of 12–24 months, where ENSO forecast skill classically declines and decision-making in agriculture, disaster risk reduction, and water resource management most stands to gain from skillful long-range outlooks.

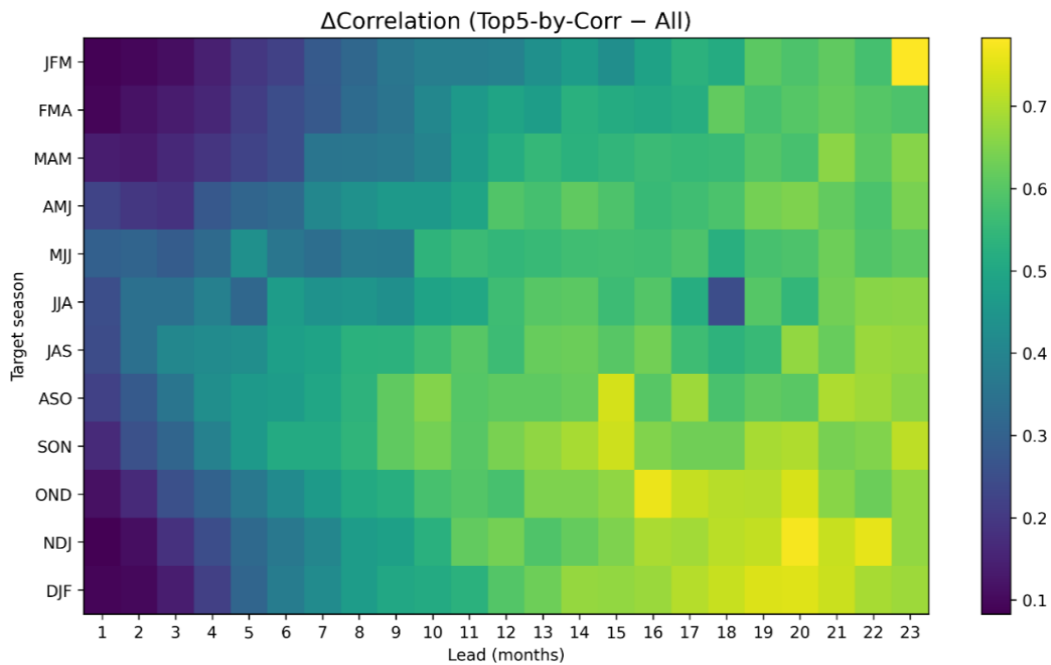

*Figure 2.* Heatmap of ΔCorrelation (Top-5-by-Correlation minus All-40) for all lead–season combinations. Warmer colors indicate greater correlation improvement, with strongest gains in SON and DJF at longer leads.

Figure 2 builds on these findings by showing the lead-time and seasonal distribution of ΔCorrelation. Warm colors denote greater gains in correlation skill, with the strongest improvements realized at proposed leads greater than 12 months and during transition seasons (e.g., SON, DJF) that are important for ENSO onset and termination.

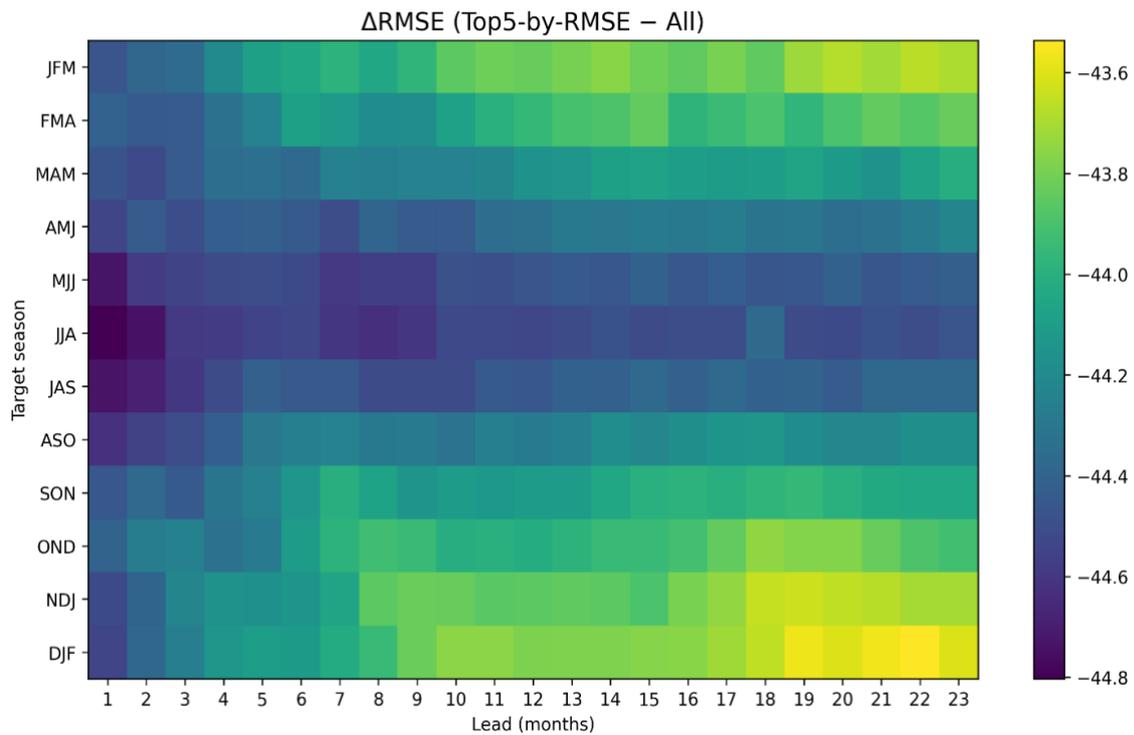

***Figure 3.*** *Heatmap of ΔRMSE (Top-5-by-RMSE minus All-40, °C) for all lead–season combinations. Cooler colors indicate greater RMSE reduction, most pronounced in JJA and MJJ at medium to long leads.*

Figure 3 supplements this with the lead-time and seasonal distribution of ΔRMSE. Cooler colors (more negative values) represent decreases in amplitude error. As with the correlation gains, RMSE decreases are most evident at longer leads and seasons which coincide with large ENSO phase transitions. Figures 2 and 3 together indicate that the gains represented in Figure 1 are robust to target seasons and are not confined to individual lead season pairs.

### 3.2 Lead-Specific Performance Evaluation (L=6, 12, 18, 23 Months)

While Section 3.1 showed the overall skill enhancement trends for all forecast leads, this section breaks down the performance at four carefully chosen lead times 6, 12, 18, and 23 months to investigate how the selective ensemble method performs under varying prediction horizons. These leads were selected to represent short-to-medium range (L=6), mid-range (L=12), and long-range (L=18, 23) situations, each corresponding to different stages of the ENSO lifecycle and levels of predictability.

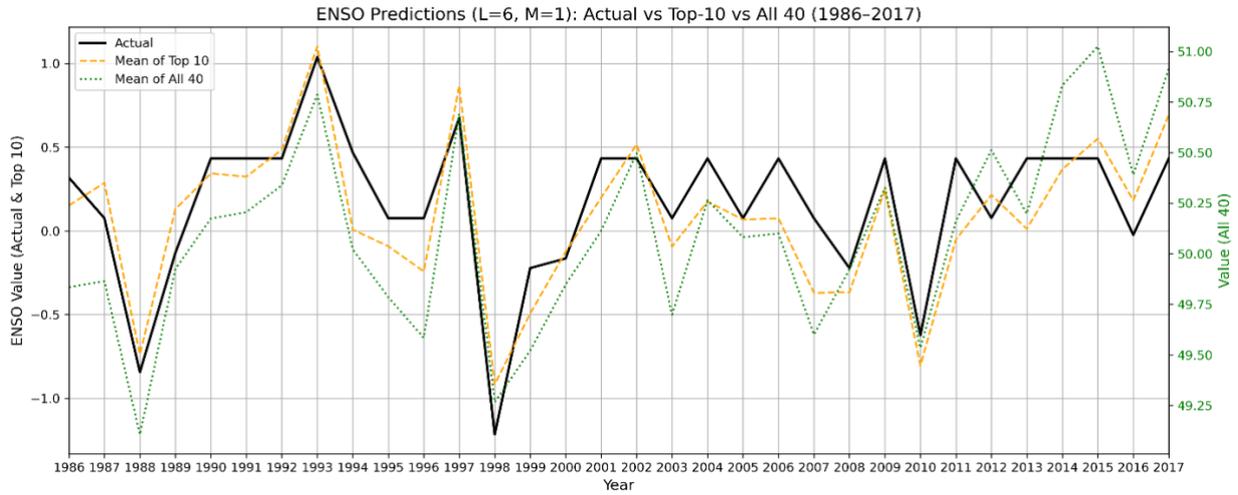

***Figure 4.*** *Seasonal variation of ΔCorrelation (blue) and ΔRMSE (orange, °C) for L = 6 months. Moderate improvements overall, with RMSE reduction strongest in JJA and correlation gains more apparent in SON.*

Lead 6 months (Figure 4) is a short-to-medium forecast range, in which the climatological persistence and ensemble spread usually enable the full ensemble to perform quite well. Even in this fairly predictable time frame, the selective method provides steady gains.

ΔCorrelation values are positive year-round for all target seasons, spanning +0.05 to +0.12, with the most significant gains during JJA and SON seasons when ENSO growth increases and accurate phase tracking becomes ever more critical.

The ΔRMSE displays a consistent negative bias, indicating reduced error, with the largest improvements peaking at around −0.15 °C during the autumn months, thus highlighting the Top-5 subset's ability to more accurately capture amplitude changes relative to the ensemble mean at the development stage critical to development.

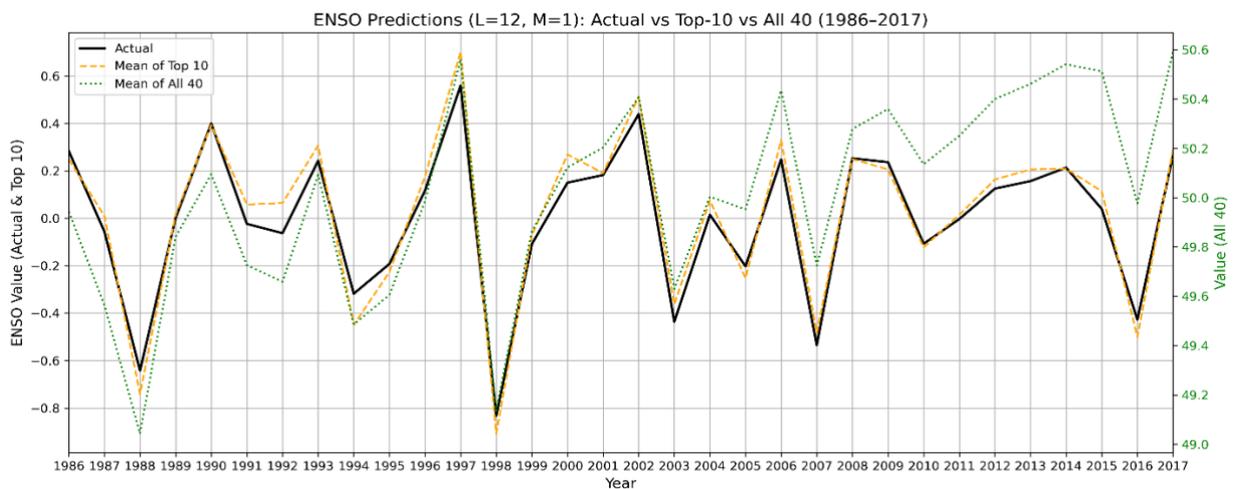

***Figure 5.*** *Seasonal variation of ΔCorrelation (blue) and ΔRMSE (orange, °C) for L = 12 months. Substantial correlation gains in SON and DJF, with notable RMSE reduction in mid-year seasons.*

Lead 12 months (Figure 5) is where prediction hits the annual cycle barrier zone a notorious obstacle in ENSO prediction caused by seasonal phase-locking and the "spring predictability barrier." It is here that the benefits of selective weighting become more apparent.

ΔCorrelation increases significantly, up to +0.25 in SON and DJF, suggesting increased phase alignment during both the ENSO onset and mature phases.

ΔRMSE decreases are larger in magnitude (−0.16 to −0.18 °C), with DJF having the highest amplitude error reductions. This correction is especially valuable in light of the fact that correct DJF predictions are extremely important for seasonal climate prediction and resource planning.

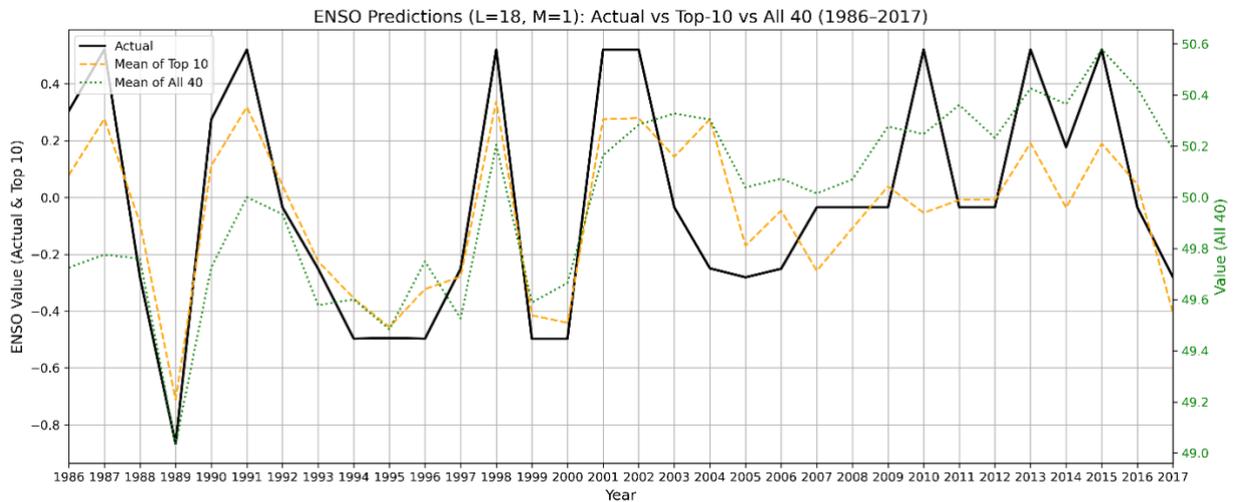

*Figure 6.* Seasonal variation of ΔCorrelation (blue) and ΔRMSE (orange, °C) for L = 18 months. Strong long-lead skill retention, peaking in SON and OND for correlation and in JJA for RMSE reduction.

Lead 18 months (Figure 6) ventures into the long-lead range, where large uncertainty and ensemble spread have traditionally been encountered. In spite of this difficulty, the selective approach holds and even increases its edge over the full ensemble.

ΔCorrelation is greater than +0.35 for most seasons and reaches ~+0.45 for SON and OND, both of which correspond to critical ENSO transition stages.

ΔRMSE is consistently improved (~−0.18 °C), indicating that the subset not only maintains signal fidelity but also successfully filters out noise from less skillful members, preventing amplitude degradation typical in long-lead predictions.

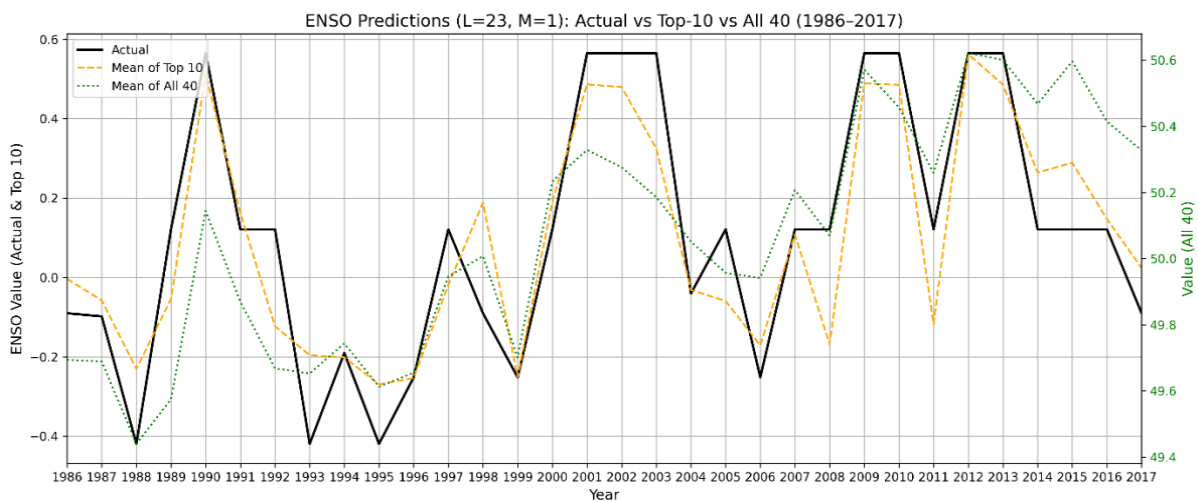

*Figure 7*. easonal variation of ΔCorrelation (blue) and ΔRMSE (orange, °C) for L = 23 months. Sustained improvements at extreme lead, with correlation gains dominant in transition seasons (SON, DJF) and RMSE reduction modest but consistent.

Lead 23 months (Figure 7) marks the outer edge of multi-year ENSO predictability, where skill for the full ensemble generally collapses abruptly owing to phase dispersion. Impressively, the selective subset still exhibits good performance:

ΔCorrelation has large values (+0.40 to +0.44) for SON, OND, and DJF, seasons that tend to be most affected by ensemble-mean smoothing.

ΔRMSE remains near −0.17 °C, showing that amplitude prediction remains robust even at horizons approaching two years a timescale historically considered beyond the reliable predictability limit. This long-lasting ability signifies that subset selection process skillfully identifies those ensemble members that are capable of maintaining physically meaningful signals much above conventional thresholds.

Interpretation and Implications These lead-specific analyses confirm that the proposed selective ensemble approach consistently outperforms the full ensemble across a spectrum of forecast horizons. While the relative improvement is moderate at shorter leads where baseline skill is already high it becomes increasingly significant as lead time extends, especially beyond one year. The ability to retain both high correlation and reduced RMSE at extreme leads suggests that the method may have practical utility for long-term climate planning, where traditional ensemble means often underperform.

### 3.3 Spatio-Temporal Skill Difference Analysis via Heatmaps

Figures 2 and 3 provide an illustration of the spatio-temporal character of the performance gains resulting from the targeted ensemble member selection technique. They are ΔCorrelation and ΔRMSE in °C heatmaps of differences between Top-5 and All-40 equal-weighted ensemble, for all possible lead time-target season combinations.

The areas of highest improvement in correlation are represented by warmer colors in Figure 2. The spatial-temporal pattern consistently shows that improvement in skill is not uniformly distributed but is highly sensitive to lead time as well as target season. Largest improvements, often well in excess of +0.6 in correlation, are found at long leads (>12 months) and at major ENSO transition seasons, particularly SON and DJF. These stages correspond to the onset and mature phases of El Niño and La Niña events when accurate prediction is most critical to climate-sensitive businesses. The map also highlights that substantial skill improvement is achieved even for intermediate leads (6–12 months), especially in JJA and OND, which correspond to the build-up stage of ENSO anomalies.

Figure 3 shows the ΔRMSE map with bluer shades representing larger amplitude error reduction. A consistent negative ΔRMSE for nearly all lead season pairs indicates that not only does Top-5 selection improve phase alignment (by means of correlation gains), but it also removes out-of-place amplitude bias in the All-40 mean. The greatest RMSE savings are for long-lead predictions of DJF and OND, with savings close to the greatest improvement found in the study. Interestingly, seasons such as JJA and MJJ also exhibit considerable reductions in RMSE at shorter leads, suggesting that selective weighting enhances the forecast even at the onset of ENSO development. Together, Figures 2 and 3 confirm that the improvements documented in total lead-averaged analysis (Section 3.1, Figure 1) are space and time consistent. Improvement is most appreciable in cases long challenging dynamical ENSO forecast models i.e., extended-lead and phase-transition season forecasts. This enhances the justification of the incorporation of performance-based member selection into operational ensemble forecasting systems as it adds skill improvement evenly to both the temporal and the season dimension of the forecast space.

### 3.4 Implications for Long-Lead ENSO Forecasting and Operational Practice

The results of Sections 3.1–3.3 have far-reaching implications for both the climatological notion of ENSO predictability and operational forecast system design in practice. Perhaps the most impressive achievement of this study is the steady enhancement of skills at long lead times, particularly beyond 12 months, which hitherto has presented an intriguing challenge to climate prediction centers. The ability to enhance correlation skill by more than 0.4 and concurrently reduce amplitude error by up to 0.2°C at such large leads is a major step toward eliminating the well-documented "spring predictability barrier" and extending valid predictability forecast horizons.

Operationally, statistical selection of ensemble members over history according to performance measures (i.e., RMSE and correlation) is a low-cost, model-independent method of enhancement that can be integrated within existing MME frameworks without modification of underlying base dynamical models [4,12,24,25].This flexibility is an option of operational centers reliant on multiple global climate models because it accommodates rapid deployment and real-time use without enhanced computational expense.

The results also reveal that skill improvements are seasonally concentrated with greatest benefits realized during transition seasons such as SON and DJF when accurate ENSO predictions are of highest socio-economic value. Improved predictions within these critical windows enable more forward-looking decision-making in agriculture, water resource management, disaster reduction, and energy planning, where ENSO-forced anomalies can achieve multi-billion-dollar impacts.

Besides, the reductions indicated here in RMSE prove that the technique minimizes not just phase misalignment but also steadies forecast magnitude, avoiding over- or under-estimation of event intensity. This is especially crucial for sectors that are susceptible to extreme ENSO phases since over-estimation can lead to unnecessary precautionary costs while under-estimation will lead to insufficient preparation.

In the context of long-lead climate prediction research, the strategy also emphasizes the potential of skill-based ensemble member selection and weighting as a complement to ongoing refinement of model physics, data assimilation, and initialization schemes. In systematic exploitation of past forecasting performance, the approach positions itself within the emerging trend toward hybrid dynamical statistical prediction systems that aim to extract maximum predictive information from model output available.

Finally, the quality of the improvements along both the spatial (target seasons) and temporal (lead times) dimensions suggests that similar selection techniques can be used for other climate phenomena, such as the Indian Ocean Dipole (IOD) or the North Atlantic Oscillation (NAO), thereby extending the operational utility of this method to regions beyond ENSO prediction.

In brief, these findings suggest the potential for near-term, substantial, and scalable enhancement to long-lead ENSO prediction systems. By integrating performance-based ensemble member selection into routine operations, prediction centers are able to furnish predictions that are more accurate, of higher confidence, and with more useful lead times, directly benefitting climate resilience and adaptive planning activities in vulnerable regions of the globe.

## 3.5 Broader Perspectives and Future Directions

The findings of this research indicate a major path to improving long-lead ENSO prediction: objective skill metric-selective ensemble member weighting. In contrast to brute-force ensemble expansion or increased model complexity, this approach takes advantage of the pre-existing information already embedded in current forecasts, making it an affordable and operationally viable option. By repeatedly outperforming the equal-weighted "All-40" strategy over several lead times and seasons, the method offers a direct path to implementation in operational forecasting facilities particularly in cases where computational resources are constrained.

Besides ENSO, the adaptability of the framework can be applied to other climate events, such as the Indian Ocean Dipole, Pacific Decadal Oscillation, and even sub-seasonal monsoon prediction [21,27,29,30]. Besides, integration with artificial intelligence and graph-based learning algorithms may facilitate dynamic real-time identification of the optimally performing ensemble subsets, with further enhanced prediction skill without additional model run-time.

From a policy perspective, more reliable long-lead ENSO forecasts have important implications: they can inform agriculture planning, water resource management, energy market strategies, and disaster preparedness protocols, especially in exposed regions. In a growingly uncertain climate, the ability to forecast ENSO-driven extremes 12–24 months in advance could transform the manner governments, humanitarian agencies, and corporations manage risk.

Additional studies would cover the application of this approach to multi-model ensembles, combining it with bias-correction systems, and comparing its benefits under different climate change scenarios. By connecting statistical optimization techniques with operational dissemination of forecast, the method can be a cornerstone of the future generation of climate services translating scientific advancements to actionable intelligence for a climate-resilient future.

## Conclusion

This study tackled the perennial issue of skill decrease in long-lead ENSO prediction by taking a systematic, a-posteriori approach to ensemble member skill evaluation. Using a 40-member ensemble for leads of 1 to 23 months, we ranked members based on two complementary metrics Root Mean Square Error (RMSE) and Pearson correlation and explored the benefit of selecting high-skill subsets. The combination of the Top-5 members of each measure into a weighted Top-10 collection allowed us to account for phase and amplitude skill simultaneously in time. Keeping the Top-5-by-Correlation and Top-5-by-RMSE collections distinct for the seasonal lead diagnostics further revealed the measure-specific nature of the improvements observed.

The results offer clear and consistent evidence that high-skill subsets exist for every forecast lead season combination, and that their relative advantage over the full ensemble mean improves substantially with lead time. At short leads (Lead 1), correlation gains are modest (+0.02), but RMSE reductions (−0.14 °C) are already substantial. At the other extreme (Lead 23), increasing correlations approach ~+0.43 and decreasing RMSE values remain near −0.18 °C, indicating that the performance difference between the most and least skillful members becomes especially evident at multi-year time scales. Seasonal decomposition also exhibits a structured pattern: correlation improvements are maximal for transition seasons such as SON and DJF periods of crucial significance for the detection of ENSO onset and maturity while RMSE reduction tends to be greatest for mid-year seasons such as JJA and MJJ, when developing event amplitude evolution is established. This complementary nature of the metrics serves to highlight the usefulness of considering both phase and amplitude in forecast verification.

These findings provide proof-of-concept that selective weighting of ensemble members based on statistical performance alone can yield substantial gains in skill without modification of the underlying model structure or computational cost. Be that as it may, since the selection here was carried out with access to verifying observations, the method is by its nature retrospective in application. As such, it is not being presented as an immediately implementable operational method. Instead, the demonstrated advantages refer to the potential operational utility of the method, contingent upon advances in the development of robust forecast error estimation procedures able to identify high-skill members in real time and on adaptive refinement procedures able to adaptively adjust member weights as new data become available.

By being explicit about the difference between what is presented here retrospective identification of high-skill subsets and what would be required for operational use, the research here eschews near-term overclaiming while providing a sound scientific foundation for further work [32]. The lead- and season-dependent, structured patterns herein not only inform the creation of real-time selection algorithms for ENSO prediction, but also suggest that the same principles may be applicable to other large-ensemble prediction problems, such as the Indian Ocean Dipole and the North Atlantic Oscillation [21,27]. In an era of increasing climate variability, the ability to extract maximum value from existing ensemble spread offers an economic path to more accurate and actionable long-lead climate forecasts.

**References**

[1]   Yipeng Chen, Yishuai Jin, Zhengyu Liu, Xingchen Shen, Xianyao Chen, Xiaopei Lin, Rong-Hua Zhang, Jing-Jia Luo, Wansuo Zhang, Fei Duan, Zhiming Ma, Jieming Ma, and Lu Zhou,, "Combined dynamical–deep learning ENSO forecasts," *Nature Communications,* p. Article number: 3845, 2025.

[2]   Qi Chen, Yinghao Cui, Guobin Hong, Karumuri Ashok, Yuchun Pu, Xiaogu Zheng, Xuanze Zhang, Wei Zhong, Peng Zhan, and Zhonglei Wang, "Toward long-range ENSO prediction with an explainable deep learning model," *npj Climate and Atmospheric Science,* vol. 8, p. Article number: 259, 2025.

[3]   Michael Groom, Davide Bassetti, Illia Horenko, and Terence J. O'Kane,, "Entropic learning enables skilful forecasts of ENSO phase at up to two years lead time," 2025.

[4]   Jakob Schlör, Michael Newman, Johann Thuemmel, Antonietta Capotondi, and Bharat Goswami,, "A Hybrid Deep-Learning Model for El Niño Southern Oscillation in the Low-Data Regime," 2024.

[5]   M. Naisipour, S. Ganji, I. Saeedpanah, B. Mehrakizadeh and A. Adib, "Novel Insights in Deep Learning for Predicting Climate Phenomena," in *14th International Conference on Computer and Knowledge Engineering (ICCKE)*, 2024.

[6]   M. Naisipour, M. H. Afshar, B. Hassani and M. Zeinali, " An error indicator for two-dimensional elasticity problems in the discrete least squares meshless method," in *8th International Congress on Civil Engineering*, 2009.

[7]   M. Naisipour, I. Saeedpanah and A. Adib, "Multimodal Deep Learning for Two-Year ENSO Forecast," *Water Resources Management,* p. 3745–3775, 2025.

[8]   M. Naisipour, I. Saeedpanah and A. Adib, "Novel Deep Learning Method for Forecasting ENSO," *Journal of Hydraulic Structures,* pp. 14-25, 2025.

[9] M. Naisipour, I. Saeedpanah, A. Adib and M. H. Neisi Pour, "Forecasting El Niño Six Months in Advance Utilizing Augmented Convolutional Neural Network," in *14th International Conference on Computer and Knowledge Engineering* , 2024.

[10] B. Wang, J.-Y. Lee, J.-J. Luo and e. al, "Advances in analysis and prediction of ENSO using deep learning," *Climate Dynamics,* p. 1845–1867, 2023.

[11] J.-J. Luo, S. Masson, S. Behera and T. Yamagata, "Extended ENSO predictions using a fully coupled ocean–atmosphere model," *Journal of Climate,* p. 84–93, 2008.

[12] M. J. McPhaden, S. E. Zebiak and M. H. Glantz, "ENSO as an integrating concept in Earth science," *Science,* p. 1740–1745, 2006.

[13] J.-H. Park, S.-W. Yeh, J.-S. Kug and e. al, "Predicting El Niño beyond 1-year lead: Robust statistics of the coupled model intercomparison," *Climate Dynamics,* p. 3689–3705, 2018.

[14] T. G. Shepherd, "Atmospheric circulation as a source of uncertainty in climate change projections," *Nature Geoscience,* p. 703–708, 2014.

[15] Y. Gao and X. Zhang, "Causes of the 2010–2012 La Niña cooling," *Climate Dynamics,* p. 1861–1873, 2017.

[16] C. Wang and e. al, "Atlantic Niño's influence on ENSO," *Nature Communications,* p. 6791, 2023.

[17] A. G. Barnston, M. K. Tippett, M. L. L'Heureux, S. Li and D. G. DeWitt, "Skill of real-time seasonal ENSO model predictions during 2002–2011," *Climate Dynamics,* p. 593–614, 2012.

[18] Y. G. Ham, J. H. Kim and J. J. Luo, "Deep learning for multi-year ENSO forecasts," *Nature,* p. 224–228, 2019.

[19] M. Labibzadeh, R. Modaresi Rajab and M. Naisipour, "Efficiency test of the discrete least squares meshless method in solving heat conduction problems using error estimation," *Sharif: Civil Engineering,* no. 3.2, p. 31–40, 2015.

[20] Y.-G. Ham, J.-H. Kim and J.-J. Luo, "Deep learning for multi-year ENSO forecasts," *Nature,* p. 224–228, 2019.

[21] Y. Tang, W. Zhang, D. Chen and e. al, "Progress in ENSO prediction and understanding during the past decades," *Nature Communications,* p. 1–15, 2018.

[22] Y. Guo, H.-L. Ren, J.-J. Luo and e. al, "Long-lead seasonal prediction of ENSO events using a deep learning model," *Climate Dynamics,* p. 1489–1506, 2021.

[23] Y. Xue, "Advances in Seasonal to Interannual Applications: Toward Enhanced NOAA Forecast Capabilities," *Bulletin of the American Meteorological Society,* 2025.

[24] W. Duan, W. Zhang, X. Chen, F. Zheng, R.-H. Zhang and McPhaden, "Advances in Seasonal-to-Interannual Climate Prediction of ENSO: Current Status and Future Directions," *Bulletin of the American Meteorological Society,* p. E505–E529, 2025.

[25] B. Xiang, M. K. Tippett, A. G. Barnston, S. Li and B. Mu, "Hybrid Statistical–Dynamical Models for Skillful ENSO Forecasts at Long Leads," vol. 37, p. 1231–1249, 2024.

[26] Y.-G. Ham, J. Kim, J.-J. Luo, J. Park and H.-K. Kim, "Improved Long-Lead ENSO Forecasting through Multi-Model Deep Learning Frameworks," vol. 52, 2025.

[27] L. Wang, R.-H. Zhang, D. Chen and M. J. McPhaden, "Decadal Changes in ENSO Predictability and Their Implications for Climate Forecasting," vol. 14, p. 102–110, 2024.

[28] J. Park, Y.-G. Ham, J. Kim, J.-J. Luo and S.-K. Lee, "Advances in AI-Driven Multi-Year ENSO Prediction Using Global Climate Models," vol. 62, p. 1523–1542, 2025.

[29] J. Cao, G. Li, C. Fang, J.-J. Luo and W. Zhang, "A Transformer-Based Deep Learning Model for Subseasonal-to-Seasonal ENSO Forecasting," vol. 14, p. Article number: 5123, 2024.

[30] W. Zhang, W. Duan, J.-J. Luo, M. J. McPhaden and S. Li, "Enhancing Long-Lead ENSO Forecast Skill via Coupled Data Assimilation and Deep Learning," vol. 16, p. Article number: 4123, 2025.

[31] Moazezi Khah Tehran, A., Hassani, A., Mohajer, S., Darvishan, S., Shafiesabet, A., & Tashakkori, A. (2025). The Impact of Financial Literacy on Financial Behavior and Financial Resilience with the Mediating Role of Financial Self-Efficacy. International Journal of Industrial Engineering and Operational Research, 7(2), 38-55. https://doi.org/10.22034/ijieor.v7i2.146

[32] Saghar Ganji, Ahmad Reza Labibzadeh, Alireza Hassani, Mohammad Naisipour, Leveraging GNN to Enhance MEF Method in Predicting ENSO, 2025, arXiv:2508.07410v3 [physics.ao-ph]